\documentclass[12pt,a4paper]{article}

\usepackage{epsfig,amsmath,amssymb}

\begin{document}

\baselineskip0.25in

\title{Adiabatic renormalization
in theories with modified dispersion relations}

\author{D. L\'opez Nacir\thanks{e-mail: dnacir@df.uba.ar},\ F. D. Mazzitelli\thanks{e-mail: fmazzi@df.uba.ar}\ \  and C. Simeone\thanks{e-mail: csimeone@df.uba.ar}\\
{\small Departamento de F\'{i}sica J. J. Giambiagi,}\\
{\small  Facultad de Ciencias Exactas y Naturales,}\\
{\small Universidad de Buenos Aires, Ciudad Universitaria, Pabell\'on 1,}\\
{\small 1428 Buenos Aires, Argentina}}

\maketitle

\begin{abstract}
We generalize the adiabatic renormalization to theories with
dispersion relations modified at energies higher than a new scale
$M_C$. We obtain explicit expressions for the mean value of the
stress tensor in the adiabatic vacuum, up to the second adiabatic
order. We show that for any dispersion relation the divergences
can be absorbed into the bare gravitational constants of the
theory. We also point out that, depending on the renormalization
prescription, the renormalized stress tensor may contain  finite
trans-Planckian corrections even in the limit
$M_C\rightarrow\infty$.
\end{abstract}



\def\ii{\'{\i}}
\def\be{\begin{equation}}
\def\ee{\end{equation}}
\def\lp{\left(}
\def\rp{\right)}
\def\lb{\left[}
\def\rb{\right]}
\def\om{\omega}
\def\La{\Lambda}
\def\la{\lambda}
\def\ck{\chi_k}

\section{Introduction}

Inflationary scenarios provide an explanation for the large scale
structure of the Universe and for the anisotropy in the Cosmic
Microwave Background (CMB). The exponential (or quasi exponential)
expansion stretches the physical wavelengths, so that a density
fluctuation which is today of cosmological scale  was originated
during inflation on scales much smaller than the Hubble radius. If
the inflationary period lasts enough to solve the causality and
other problems, the scales of interest today are not only within
the horizon but are also {\it sub-Planckian} at the beginning of
inflation \cite{bran1}. This fact, known as the {\it
trans-Planckian problem}, provides a potential window to observe
consequences of the Planck scale physics. Hence the possibility of
observing signatures of Planckian physics in the power spectrum of
the CMB and in the evolution of the Universe has been widely
studied. In the absence of a full quantum theory of gravity, the
analysis must be  phenomenological. One possibility is to consider
an effective field theory approach in which the new physics is
encoded in the state of the quantum fields when they leave
sub-Planckian scales \cite{eff}. Other possibility, which will be
analyzed here, is to consider modified dispersion relations for
the modes of quantum fields, which might arise in loop quantum
gravity or due to the interaction with gravitons \cite{grav}. It
is important to test the robustness of inflationary predictions
under such modifications.

In simple models with a  scalar field $\phi$, the information on
the power spectrum of the CMB is contained in the two point
function $\langle \phi (x)\phi (x')\rangle$, and the backreaction
of the field on the spacetime metric is contained in the
expectation value $\langle T_{\mu\nu} \rangle$. A consistent
treatment of the backreaction problem should rely in a careful
evaluation of the expectation value of the energy density and
pressure, acting as the source in the Semiclassical Einstein
Equations (SEE).  In general, $\langle T_{\mu\nu}\rangle$ is a
divergent quantity. In previous works \cite{branma,lemoine}, the
renormalization prescription consisted basically in subtracting
the ground state energy of each Fourier mode, but this may lead to
inconsistencies for quantum fields in curved spaces \cite{wald}.

The renormalization procedure for quantum fields satisfying the
standard dispersion relations in curved backgrounds is well
established \cite{wald}. The adiabatic renormalization
\cite{adiab}, consists in the subtraction of the stress tensor
constructed with the WKB expansion of the field modes, up to the
fourth adiabatic order. The divergences are proportional to
geometric conserved tensors, and can be absorbed into the bare
constants of the theory. Thus one defines the renormalized stress
tensor as $\langle T^{\mu\nu}\rangle-\langle
T^{\mu\nu}\rangle^{({\mathrm {Ad}})}$. Here we show that in the
case of generalized dispersion relations $\omega^2\sim k^{2r},\
r\geq 2$, fourth or higher adiabatic order contributions are
already finite in $3+1$ dimensions. This suggests that no terms
quadratic in the curvature would be necessary in the SEE, and only
a redefinition of the cosmological constant and the Newton
constant would be required to absorb the divergent contributions.
However, as we will see, this naive argument is incorrect: this
prescription would lead to a discontinuity in the order of the
adiabatic subtraction, and  may leave a mark of trans-Planckian
physics as a non vanishing contribution to the stress tensor even
in the limit $M_C\rightarrow \infty$. We will discuss this
subtlety using as example the calculation of the trace anomaly in
1+1 dimensions.

\section{The WKB expansion}

We  consider  a scalar field $\phi$ with a non standard dispersion
relation induced by higher spatial derivatives  \be \om^2_k=
k^2+C(\eta)\lb
m^2+2\sum_{s,p}(-1)^{s+p}\,b_{sp}\,\lp\frac{k}{C^{1/2}(\eta)}\rp^{2(s+p)}\rb,
\label{dis} \ee where  $b_{sp}$ are arbitrary coefficients of
order $M_C^{2(1-s-p)}$, and $p\leq s$. We work with a general
spatially flat FRW metric given by $
ds^2=C(\eta)[-d\eta^2+\delta_{i j}dx^i dx^j] $ where
$C^{1/2}(\eta)$ is the scale factor given as a function of the
conformal time $\eta$.

The Fourier modes $\chi_k$ corresponding to the scaled field
$\chi=C^{(n-2)/4}\phi$ satisfy \be
\chi_k''+\lb(\xi-\xi_n)RC+\om_k^2\rb\chi_k=0\, . \label{PXXP}\ee
Here primes stand for derivatives with respect to the conformal
time $\eta$, $R$ is the Ricci scalar, and $\xi$ defines the
coupling with the curvature. The field modes $\chi_k$ can be
expressed in the well known form \be \ck= \frac{1}{\sqrt{ 2
W_k}}\exp\lp -i\int^\eta W_k(\tilde\eta)d\tilde\eta\rp
.\label{chi} \ee Substitution of Eq. (\ref{chi}) into Eq.
(\ref{PXXP}) yields a nonlinear differential equation for $W_k$
that can be solved iteratively by assuming that $W_k$ is a slowly
varying function of $\eta$. In this WKB approximation, the
adiabatic order of a given term is defined as the number of
derivatives of the metric. Working up to the second adiabatic
order, we straightforwardly obtain \cite{lms}
\begin{eqnarray}
W_k^2 & = & \om_k^2+\lp \xi-\xi_n\rp(n-1)\lp\frac{C''}{C}+\frac{(n-6)}{4}\frac{{C'}^2}{C^2}\rp -
\frac{1}{4}\frac{C''}{C}\lp 1-\frac{k^2}{\om_k^2}\frac{d\om_k^2}{d k^2}\rp\nonumber\\
 &-& \  \frac{1}{4}\frac{{C'}^2}{C^2}\frac{k^4}{\om_k^2}\frac{d^2\om_k^2}{{d(k^2)}^2}
 +   \frac{5}{16}\frac{{C'}^2}{
C^2}\lp1-\frac{k^2}{\om_k^2}\frac{d\om_k^2}{d k^2}
\rp^2\label{W2}.
\end{eqnarray}
From Eq. (\ref{W2}) it is clear that while the zeroth adiabatic
order scales as $\om_k^2$, the second order scales as $\om_k^0$.
It can be shown that the $2j-$adiabatic order scales as
$\om_k^{2-2j}$.

\section{Renormalization of the stress tensor}

We start from the vacuum expectation values of the energy density
$\rho$ and pressure $p$, generalized to arbitrary dimension $n$
and coupling $\xi$ \cite{lms}:
\begin{eqnarray}
\nonumber\langle\rho\rangle &=& \frac{1}{ \sqrt{C}}\int
\frac{d^{n-1}k\,\mu^{\bar n - n}}{(2\pi\sqrt{C})^{(n-1)}} \left\{
\frac{C^{(n-2)/2}}{2}\left|\lp \frac{\chi_k}{C^{(n-2)/4}}\rp'
\right|^2 +\xi G_{\eta\eta}|\chi_k|^2\right.
\\
 &+&\left.\frac{\om_k^2}{2}\,|\chi_k|^2+\xi\frac{(n-1)}{2}
\lb\frac{C'}{C}(\chi_k'\chi_k^*+\chi_k{\chi_k'}^{*})-\frac{{C'}^2}{C^2}\frac{(n-2)}{2}|\chi_k|^2
\rb \right\},\label{RHOO}\\
\nonumber\langle p\rangle &=& \frac{1}{ \sqrt{C}}\int
\frac{d^{n-1}k\,\mu^{\bar n-n}}{(2\pi\sqrt{C})^{(n-1)}}
\left\{\lp\frac{1}{2}-2\xi\rp C^{(n-2)/2}\left|\lp
\frac{\chi_k}{C^{(n-2)/4}}\rp' \right|^2\right.\\ \nonumber
 &+&\xi
G_{11}|\chi_k|^2+\lb \lp\frac{k^2}{n-1}\rp\frac{d\om_k^2}{
dk^2}-\frac{\om_k^2}{2}\rb|\chi_k|^2-\xi(\chi_k''\chi_k^*+\chi_k{\chi_k''}^{*})\\
&+&\left.\xi\frac{
C'}{2C}(\chi_k'\chi_k^*+\chi_k{\chi_k'}^{*})+\xi\frac{(n-2)}{2}\lp\frac{C''}{C}-\frac{(8-n)}{4}\frac{{C'}^2}{C^2}\rp|\chi_k|^2\right\}.\label{PPP}
\end{eqnarray}
Here $\bar n$ is the number of dimensions of the physical
spacetime, $\mu$ is an arbitrary parameter with mass dimension and
$G_{\eta\eta}$ and $G_{11}$($=G_{22}=...=G_{n-1,n-1}$) are
the nontrivial components of the Einstein tensor.

The vacuum expectation values $\langle\rho\rangle$ and $\langle
p\rangle$ are found from Eqs. (\ref{chi}), (\ref{RHOO}) and
(\ref{PPP}). Knowing the dependence with $k$ of the $2j-$adiabatic
order one can show that, for $\om_k^2\sim k^{2r}$ with $r\geq 4$,
all contributions of second or higher adiabatic order are finite.
The divergences come only from the zeroth order terms contained in
$\langle\rho\rangle$ and $\langle p\rangle$. Instead, in the cases
$\om_k^2\sim k^6$ and $\om_k^2\sim k^4$, though no fourth order
divergences appear, second  order terms include divergent
contributions in $3+1$ dimensions.

The zeroth and second adiabatic orders of the stress tensor can be
computed from  Eqs. (\ref{chi}) to (\ref{PPP}).  After a long
calculation we obtain \cite{lms} \begin{equation} \langle
T_{\mu\nu}\rangle^{(0)} = - \frac{g_{\mu\nu}}{
4}\frac{\Omega_{n-1}\,\mu^{\bar n-n}}{(2\pi)^{n-1}} \int_0^\infty dx\,x^{(n-3)/2}\tilde{\om}_k ,
\label{ZeO}\end{equation}
\begin{equation}
\langle T_{\mu\nu}\rangle^{(2)}  =
G_{\mu\nu}\frac{\Omega_{n-1}\,\mu^{\bar n
-n}}{4\,(2\pi)^{n-1}}\left\{
  \frac{I_2}{6(n-1)(n-2)}+(\xi - \frac{1}{6}) I_1\right\},\label{rhoxi}  \end{equation}
\begin{equation} I_1  =   \int_0^\infty dx\,\frac{x^{(n-3)/2}}{\tilde{\om}_k},\ \ \ \ \ \ \ \ \ \
I_2 = \int_0^\infty
dx\,\frac{x^{(n+1)/2}}{\tilde{\om}_k^3}\frac{d^2\tilde{\om}_k^2}{{d
x}^2}\label{integrals}, 
\end{equation}
where $\Omega_{n-1}\equiv 2\pi^{(n-1)/2}/\Gamma[(n-1)/2]$,
$x\equiv k^2/C$ and $\tilde{\om}_k=\om_k/\sqrt{C}$. To get these
results, we performed several integrations by parts, and used the
fact that in dimensional regularization the integral of a total
derivative vanishes. For the usual dispersion relation, the
integral $I_2$ vanishes and we recover the known second adiabatic
order results \cite{adiab}.

Eqs. (\ref{ZeO}), (\ref{rhoxi}) and (\ref{integrals}) show that $
\langle T_{\mu\nu}\rangle^{(0)} =  N_0 g_{\mu\nu} $ and $\langle
T_{\mu\nu}\rangle^{(2)}  =  N_2 G_{\mu\nu}, $ where $N_0$ and
$N_2$ are divergent factors in $3+1$ dimensions. Hence these
contributions can be absorbed into a renormalization of the
cosmological and Newton constants: we can define $\langle
T_{\mu\nu}\rangle_{\mathrm{Ren}}=\langle T_{\mu\nu}\rangle-
\langle T_{\mu\nu}\rangle^{(0)}-\langle T_{\mu\nu}\rangle^{(2)} $
and write the SEE as \be G_{\mu\nu}+\Lambda_{\mathrm
R}g_{\mu\nu}=8\pi G_{\mathrm R}\langle
T_{\mu\nu}\rangle_{\mathrm{Ren}}. \ee Differing from the standard
case $\om_k^2\sim k^2$, all contributions of adiabatic orders
higher than the second are finite, which suggests that no terms
quadratic in the curvature would be necessary in the SEE.

\section{A discontinuity in the limit $M_C\rightarrow\infty$ ?}

Assuming that the usual dispersion relation is valid for all
energies, the renormalization of the stress tensor in 3+1
dimensions is achieved by subtracting the fourth adiabatic order
\cite{wald,adiab}. This procedure is equivalent to a redefinition
of the bare constants in the effective Lagrangian, and is
independent of the particular metric considered. It is necessary
to renormalize not only the cosmological and Newton constants, but
also to include in the gravitational Lagrangian the three
quadratic terms $\alpha_0 R^2 +\beta_0 R_{\mu\nu}R^{\mu\nu}
+\gamma_0 R_{\mu\nu\rho\sigma} R^{\mu\nu\rho\sigma}$, and absorb
the infinities into the bare constants $\alpha_0$, $\beta_0$ and
$\gamma_0$. The trace anomaly is a well known consequence of
this renormalization scheme. However, if at some very high energy
scale $M_C$ the dispersion relation gets modified by a term of the
form $k^{2r}/M_C^{2(r-1)}$, $r\geq 2$, only the second adiabatic
order is divergent and, as described above, it would be enough to
``dress'' the cosmological and Newton constants. This
``discontinuity'' in the order of the subtraction may produce non
vanishing trans-Planckian contributions to the renormalized stress
tensor, even in the limit $M_C\rightarrow\infty$.

However, there is a subtle point in this argument. While by power
counting the fourth adiabatic order is finite, divergences may
appear when $\langle T_{\mu\nu}\rangle^{(4)}$ is expressed in
terms of geometric tensors in $n$ dimensions. Indeed, this is the
case in lower dimensions, as we will illustrate in what follows.
Let us consider a conformally coupled ($\xi=0$) massless field in
$1+1$ dimensions, with a dispersion relation of the form
$\omega^2=k^2+2bk^4/C(\eta)$. For FRW metrics in $n$ dimensions
$G_{\mu\nu}$ is proportional to $n-2$ (this tensor vanishes as
$n\rightarrow 2$, because it is the variation of the would be
Gauss-Bonnet topological invariant at $n=2$). Therefore, from Eqs.
(\ref{ZeO}), (\ref{rhoxi}) and (\ref{integrals}) one readily sees
that only the zeroth adiabatic order is divergent in the limit
$n\rightarrow 2$. One would naively conclude that the subtraction
of this order would be enough, even for the usual dispersion
relation, where it is known that it is necessary to subtract up to
the second adiabatic order \cite{wald}. The point is that while
the geometric tensor $G_{\mu\nu}$ vanishes in exactly two
dimensions, in Eq. (\ref{rhoxi}) it is multiplied by a function
that has a simple pole at $n=2$. Therefore, as the poles must be
absorbed into the bare constants  before taking the limit
$n\rightarrow 2$, it is necessary to subtract up to the second
adiabatic order, whatever the dispersion relation. The
renormalized trace of the stress tensor is then defined as
\begin{equation}
\langle T \rangle_{\mathrm{Ren}} = \langle T\rangle_{\mathrm{modes}} -\langle
T\rangle^{(0)}- \langle T\rangle^{(2)} \label{trace2}
\end{equation}
where $\langle T\rangle_{\mathrm{modes}}$ is the unrenormalized
trace computed using Eq. (\ref{PPP}) and the explicit expressions
for the modes of the field. The case $b=0$ is very well known:
$\langle T\rangle_{\mathrm{modes}}$ vanishes due to conformal
invariance, $\langle T\rangle^{(0)}$ also vanishes and one
recovers the usual trace anomaly $\langle
T\rangle_{\mathrm{Ren}}=-\langle T\rangle^{(2)}=R/24\pi$.

Let us now assume that $b\neq 0$, and compute the renormalized
trace in de Sitter space $C(\eta)=\alpha^2/\eta^2$. We have
\begin{equation}
\langle T\rangle_{\mathrm{modes}}- \langle T\rangle^{(0)} =
-\frac{1}{\pi C} \int_0^\infty dk
\left(1-\frac{k^2}{\om_k^2}\frac{d\om_k^2}{dk^2}\right)\left(\om_k^2\vert
\chi_k\vert^2 -\frac{\om_k}{2}\right)
\end{equation}
The modes of the field satisfy
\begin{equation}
|\chi_k|^2=\frac{e^{-\lambda\pi/4}}{k}\sqrt{\frac{\lambda}{2}}|D_{-(1+i\lambda)/2}[(i+1)s]|^2\equiv
\frac{1}{k}f(\lambda, s), \end{equation} where $D$ is the
parabolic function, $s=k\eta/\sqrt{\lambda}$ and
$\lambda=\alpha/\sqrt{2 b}$. After changing variables and some
algebra we get
\begin{equation} \langle T\rangle_{\mathrm{modes}}- \langle T\rangle^{(0)} = \frac{R}{2\pi }\int_0^{\infty} ds
s^3\left\{f(\lambda,s)-\frac{\sqrt{\lambda}}{2\sqrt{\lambda+s^2}}\right\}\label{tm}
\end{equation}
A numerical evaluation of the integral gives, in the limit
$M_C\rightarrow\infty$, $\langle T\rangle_{\mathrm{modes}}-
\langle T\rangle^{(0)}=-R/24\pi$. As for the case of the usual
dispersion relation $b=0$, the trace of the stress tensor has an
anomaly. However, the numerical value does not coincide with the
usual one (it differs by a sign). Therefore, if we subtract only
the zeroth adiabatic order, there is a discontinuity in the
renormalized stress tensor as $M_C\rightarrow\infty$. This
discontinuity disappears if, as already mentioned, we also
subtract the second adiabatic order. Indeed, from Eq.
(\ref{rhoxi}) we find, near $n=2$,
\begin{equation}
\langle
T\rangle^{(2)}=-\frac{R}{48\pi}\left((2-n)I_1+I_2\right)\mu^{2-n}\label{t2}
\end{equation}
As $I_1$ is finite for nonvanishing $b$, the first term does not
contribute to the trace in $n=2$. On the other hand, an explicit
evaluation of $I_2$ gives,  in the limit $b\rightarrow 0$,
$\langle T\rangle^{(2)}=-R/12\pi$. Therefore, combining Eqs.
(\ref{trace2}), (\ref{tm}) and (\ref{t2}) we see that the usual
trace anomaly is recovered in the limit $M_C\rightarrow\infty$.

\section{Conclusions}

The adiabatic subtraction can be generalized to theories with
modified dispersion relations: for any dispersion relation, the
zeroth and second adiabatic orders of the stress tensor are
proportional to $g_{\mu\nu}$ and $G_{\mu\nu}$ respectively. In
$3+1$ dimensions, the higher powers of $k^2$ in the dispersion
relation make finite the fourth adiabatic order. Therefore, in
order to get a finite mean value of the stress tensor, it would be
enough to subtract up to the second adiabatic order. This would be
possible for any value of the new physics scale $M_C$. However, as
we have shown with a simple example in $1+1$ dimensions, this
renormalization prescription, which is not equivalent to a
redefinition of the bare constants in the effective Lagrangian of
the theory, would lead to a discontinuity in the stress tensor as
$M_C\rightarrow\infty$, and to a wrong value of the trace anomaly.
It is likely that the finite fourth adiabatic order near 3+1
dimensions, when written in terms of $n$-dimensional geometric
tensors, will also have poles. This may happen for a term
proportional to the variation of the would be Gauss-Bonnet
topological invariant near $n=4$. In this situation, a consistent
renormalization would involve the subtraction of the fourth
adiabatic order. Work in this direction is in progress.

\vskip.8cm

\noindent This work has been supported by  Universidad de Buenos Aires,
CONICET and ANPCyT.

\end{document}